\documentclass[final,1p,11pt]{elsarticle}

\usepackage{fourier}

\usepackage{amssymb}

\usepackage{amsmath,amsopn}

\usepackage{booktabs,xspace,bm}

\usepackage{microtype}

\journal{Journal of Computational Physics}

\begin{document}

\begin{frontmatter}



\title{A note on the carbuncle in shallow water simulations}


\author{Friedemann Kemm}

\ead{kemm@math.tu-cottbus.de}

\address{Institute for Applied Mathematics and Scientific Computing,
    Brandenburgische Technische Universit\"at Cottbus,
    Platz der Deutschen Einheit 1, 03046 Cottbus, Germany}

\begin{abstract}
An important problem in gas dynamics simulation is to prevent the
carbuncle, a breakdown of discrete shock profiles. We show that for
high Froude number, this also occurs in shallow water simulations and
give numerical evidence that all cures developed for gas dynamics
should also work in shallow water flows. 
\end{abstract}

\begin{keyword}
Shallow Water \sep Carbuncle \sep High Froude Number


    \MSC[2010] 35L65 
    \sep 65M08 
    \sep 76L05 
    \sep 76M12 

\end{keyword}

\end{frontmatter}



\section{Introduction}
\label{sec:int}

Starting with the seminal paper by Quirk~\cite{quirk}, many
investigations of the carbuncle, the breakdown of discrete shock
profiles, in gas dynamics and many strategies to suppress it were
published,
e.\,g.~\cite{sanders98,bultelle98,aroracarb,chauvat05,kim03,pandolfi,park-kwon}.Among
the classical complete Riemann solvers, i.\,e.\ Riemann solvers which
explicitly resolve all waves in the Riemann problem, the Roe solver is
known to be most prone to that numerical artifact and the Osher solver
to be least prone to it. Solvers like HLLE, which does not resolve
entropy and shear waves, are known to avoid that effect completely.

The carbuncle is known to be mainly driven by an interplay between
one-dimensional shock instabilities like postshock
oscillations~\cite{aroracarb} or jumping shock
positions~\cite{bultelle98} and undamped unphysical vortices along the
shock front.

So, there are two major ways to prevent a scheme from producing
carbuncle like structures: Adjust the viscosity on strong shocks to
stabilize them (like Osher) or adjust the viscosity on entropy and
shear waves (like HLLE). Since the Osher scheme cannot completely
avoid the carbuncle~\cite{michael-carbuncle}, most cures follow the
second way, which in turn offers two major approaches: raise the
viscosity on linear waves whenever a strong shock in a neighbouring
Riemann problem is detected or raise the viscosity when the residual
in the Rankine Hugoniot condition for that wave is large,
e.\,g.~\cite{kemm-lyon}. The price for the first approach is a
nonlocal data treatment, the price for the second is a slight increase
of shear viscosity outside of single shear waves.

Since mathematically there are some similarities between the Euler
equations of gas dynamics and the shallow water
equations~\cite{toro,toro-shallow}, and since Denlinger and
O'Connell~\cite{denlinger2011closure} report carbuncle-like
instabilities in their simulations, we expect the carbuncle to also
appear in shallow water flows and to follow the same mechanisms as in
the gas dynamics case.

In the following section we will give numerical evidence of that. We
translate two standard tests from Euler to shallow water and show how
they produce the carbuncle. We also show that our simple carbuncle
cure~\cite{kemm-lyon} has the same effect in shallow water as it has
in gas dynamics. From this, we conclude that the same should be true
for all other carbuncle cures which were developed for the Euler
equations. 

\section{Numerical Investigation}
\label{sec:numer-invest}

In this section, we provide a numerical investigation of the carbuncle
in shallow water and the effect of a carbuncle cure originally
developed for gas dynamics. For the sake of simplicity, we resort to
our own simple carbuncle cure, the HLLEMCC-solver~\cite{kemm-lyon},
which is designed to prevent the carbuncle while, like Roe, yielding
exact resolution of steady shear waves. This can be literally
translated to shallow water. The parameters given
in~\cite[Section~4]{kemm-lyon} are used with one difference: we can
use~\(\varepsilon = 10^{-5}\) instead of~\(\varepsilon = 10^{-2}\). We
compare this to the Roe and to the HLLE solver. The computational
domain for all test cases is~\([-2.5,2.5]\) discretized
with~\(40\times 40\) grid cells. The graviational constant is
normalized to one. The computations are done in
clawpack~\cite{clawpack} with first order and flat bed.


\subsection{Colliding flow test}
\label{sec:colliding-flow-test}

LeVeque~\cite[Section~7.7]{astro-leveque} outlined a test problem for
the carbuncle instability. It consists of a pair of slowly moving
shocks, initialised by a strong colliding flow. To trigger the
carbuncle, LeVeque disturbs the initial state in one grid point. In
this study, we simulate this situation for shallow water by setting
the initial water-height to~\(h=1\) and the left and right velocities
to~\(u=\pm 30\). The transverse velocity is zero. Onto this initial
state, we superimpose artificial numerical noise of amplitude
\(10^{-6}\) instead of disturbing it in just one point. The noise is
generated randomly. This allows us to make sure that the resulting
structure of the solution is independent of the structure of the
initial perturbation.
\begin{figure}
  \centering
  \includegraphics[width=.7\linewidth]{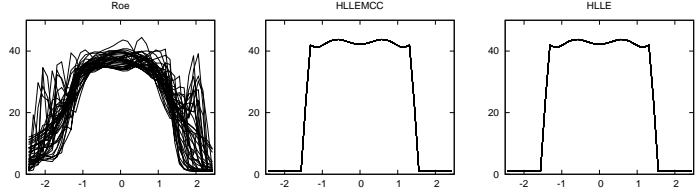}
  \caption{Scatter plot of colliding flow at time \(t=2\) with Roe,
    HLLEMCC, HLLE (left to right), water height shown.}
  \label{fig:colliding}
\end{figure}
Figure~\ref{fig:colliding} shows that, while the results for HLLE and
HLLEMCC are indistinguishable, the Roe scheme produces carbuncle type
structures.

\subsection{Steady shock test}
\label{sec:steady-shock-test}

The steady shock problem was introduced by Dumbser
et~al.~\cite{michael-carbuncle} as a test for the carbuncle. We
consider the worst case: The shock is located directly on a cell face.
According to Dumbser et~al.~\cite{michael-carbuncle} this situation is
most likely to evolve a carbuncle-like structure. For shallow water we
set the water height and the Froude number at the inflow to~\(h=1\)
and \(Fr=30\). Again, we add artificial numerical noise, this time of
amplitude \(10^{-3}\), to the conserved variables in the initial
state.
\begin{figure}
  \centering
  \includegraphics[width=.7\linewidth]{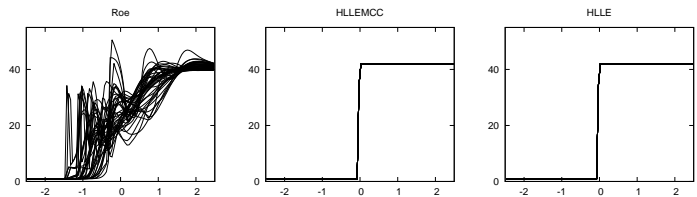}
  \caption{Scatter plot of steady shock with inflow Froude number \(Fr=30\) at time
    \(t=2\) with Roe, HLLEMCC, HLLE (left to right), water height shown.}
  \label{fig:steady}
\end{figure}
Figure~\ref{fig:steady} shows results for the water height at
time~\(t=2\). Again, the Roe scheme shows a breakdown of the shock
profile, while HLLE and HLLEMCC nicely reproduce the shock. 

\begin{figure}
  \centering
  \includegraphics[width=.7\linewidth]{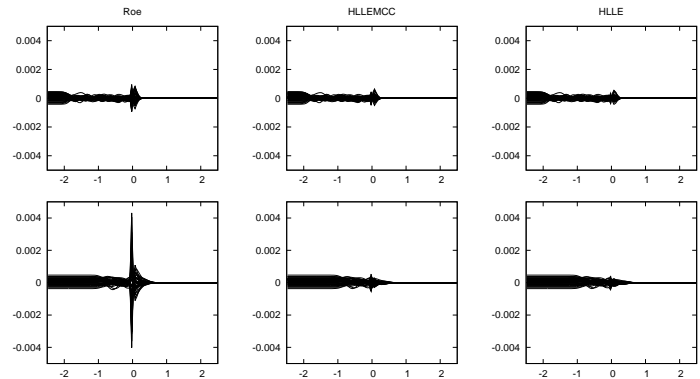}
  \caption{Scatter plot of transverse velocity component (\(v\)) at times \(t=0.002\) (upper row) and \(t=0.006\) (lower row) with Roe, HLLEMCC, HLLE (left to right).}
  \label{fig:rising}
\end{figure}

To illustrate the aforementioned interplay between 1d-instabilities of
the shock and shear flows and vortices along the shock front, in
Figure~\ref{fig:rising}, we plot the transverse velocity. 
It is nicely seen that the Roe scheme leaves the shear flow along the
shock undamped and, thus, let it grow until it starts to interact with
the shock front and destroy it. Since along the strong shock the
residual in the Rankine Hugoniot condition for the shear wave is
rather large, HLLEMCC puts almost the same amount of viscosity on it
as with the HLLE. The shear flow is damped and, thus, the resulting
vortices are to weak to act back on the shock profile.

\section{Conclusions and Outlook}
\label{sec:conclusions-outlook}

We have given numerical evidence that the carbuncle in shallow water
not only exists, but has even the same nature as in the gas dynamics
case. It is driven by a complex interplay between one-dimensional
shock instabilities and undamped unphysical vortices along the shock
front. And it can be cured by a modified diffusion mechanism which was
originally developed for the Euler equations. Thus, we expect that any
carbuncle cure developed for the Euler equations should also work for
shallow water flows.




\bibliographystyle{elsarticle-num} 
\bibliography{shcarb}






\end{document}